# Dynamics of charge states at the surface of a ferroelectric nanoparticle in a liquid crystal

Juliya M. Gudenko[1], Oleksandr S. Pylypchuk[1], Victor V. Vainberg[1,*], Vladimir N. Poroshin[1], Denis O. Stetsenko[1], Igor A. Gvozdovskyy[1], Oleksii V. Bereznykov[1], Serhii E. Ivanchenko[2], Eugene A. Eliseev[2,†], Anna N. Morozovska[1,‡]

[1]*Institute of Physics of the National Academy of Sciences of Ukraine, 46, Nauky Ave., 03028 Kyiv, Ukraine*

[2]*Frantsevich Institute of Materials Science, the National Academy of Sciences of Ukraine, 3, Omeliana Pritsaka, 03142 Kyiv, Ukraine*

## Abstract

The liquid crystal with suspended ferroelectric nanoparticles is an interesting object for fundamental research of the long-range dipole-dipole interactions; as well as it is promising for optical, optoelectronic and electrochemical applications. Such suspensions can serve as basic elements for advanced nonvolatile memory cells and energy storage devices. The work studies the cells filled with a nematic liquid crystal 5CB and the cells containing 5CB with 0.5 wt.% and 1 wt.% of $BaTiO_3$ nanoparticles with an average size of 24 nm. We analyzed the time dependences of the current flowing through the cells at constant applied voltage and the voltage dynamics in the no-load mode. The time dependences of the current and voltage show a slowing down decay rate. For the cells with $BaTiO_3$ nanoparticles, the decrease in the decay time is characteristic. A possible physical reason for the retarding decay time rate is the indirect effect of screening charges, which cover ferroelectric nanoparticles, and slow ionic transport in the liquid crystal. To explain the dynamics of current and voltage, the finite element modeling of the polarization distribution, domain structure dynamics, and charge state of nanoparticles in a liquid crystal is performed using Landau-Ginzburg-Devonshire approach. Theoretical results confirmed the leading role of screening charges, because the surface of a ferroelectric nanoparticle adsorbs an ionic-electronic charge that partially screens its spontaneous polarization in single-domain and/or poly-domain states. When an electric field is applied to the liquid crystal with nanoparticles, it can release part of the screening charge (mainly due to the change in the polarization of the nanoparticle), which will lead to a decrease in decay time rate of the current and voltage dependences.

---

[*]corresponding author, e-mail:viktor.vainberg@gmail.com
[†]corresponding author, e-mail:eugene.a.eliseev@gmail.com
[‡]corresponding author, e-mail:anna.n.morozovska@gmail.com

## 1. Introduction

Studies of the unique electrophysical properties of ferroelectric nanoparticles, the sizes of which vary from 5 to 50 nm, are of considerable fundamental and practical interest [1, 2, 3, 4, 5]. It is worth highlighting several studies of ferroelectric $BaTiO_3$ nanoparticles covered by heptane and oleic acid, whose structure becomes a core-shell type structure [6, 7, 8, 9, 10]. In these studies, the nanoparticles suspended in a liquid dielectric medium, unexpectedly have a "giant" spontaneous polarization, which reaches 1.3 C/m$^2$ at room temperature. This value exceeds the spontaneous polarization of a $BaTiO_3$ single crystal in more than 5 times. Core-shell nanoparticles with "giant" spontaneous polarization can significantly enhance the electrocaloric response [11] of colloids, and influence the phase transitions dynamics in nematic liquid crystals [6-10].

A significant enhancement of spontaneous polarization in small $BaTiO_3$ nanoparticles can be caused both by the strain gradient and/or the curvature of their surface (since the ferroelectric is very sensitive to various types of deformations), and by the electrochemical interaction of the particle core with the shell and/or the surrounding environment. In particular, it has been demonstrated experimentally [12], that the spontaneous polarization $P_S$ of $BaTiO_3$ films increases by 250% under epitaxial strain (lattice mismatch), while the Curie temperature $T_C$ increases to approximately 500°C. First-principles calculations [13] and thermodynamic theory [14] explain the experimental results. It was subsequently shown that electrochemical strains are responsible for the strong increase in the Curie temperature (above 167 °C) and tetragonality (up to 1.032) near the surface of BaTiO$_3$ films with oxygen vacancies [15]. The lattice constant mismatch, which causes epitaxial and/or core-shell strains, is responsible for the enhancement of $P_S$, $T_C$, and tetragonality in the core-shell BaTiO$_3$ nanoparticles [16, 17, 18].

The issue of the electrochemical interaction of the nanoparticle core with its shell and/or the surrounding environment remains largely unexplored. In this context research of electrophysical properties of small BaTiO$_3$ nanoparticles suspended in a liquid crystal is of significant scientific interest. The liquid crystal with suspended BaTiO$_3$ nanoparticles is promising for optical, optoelectronic and electrochemical applications. Such suspensions can serve as basic elements for advanced nonvolatile memory cells and energy storage devices [19].

## 2. Analysis of experimental results

We prepared the cells filled with pure nematic liquid crystal 5CB and liquid crystal suspension containing 0.5 wt.% and 1 wt.% of BaTiO$_3$ nanoparticles with an average size of 24 nm. A schematic illustration of the cells preparation is presented in **Fig. 1(a)**. A schematic illustration of a BaTiO$_3$ nanoparticle, which is covered with a shell of screening ionic-electronic charge and surrounded with elongated polar molecules of the liquid crystal, is presented in **Fig. 1(b)**.

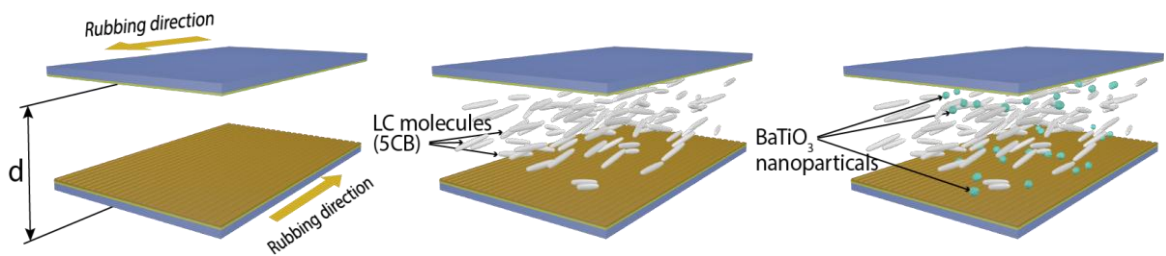

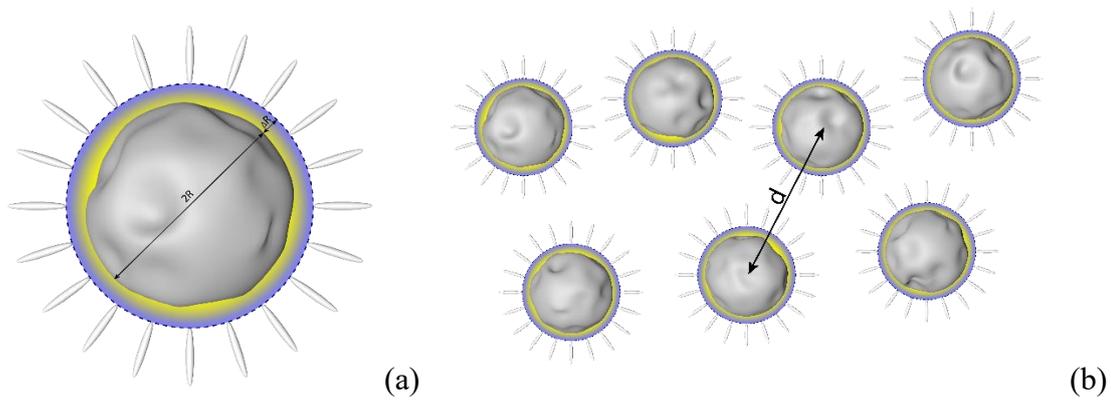

**Fig. 1. (a)** A BaTiO$_3$ core with an average size 2R covered by a screening shell of ionic-electronic charge. **(b)** The core-shell nanoparticles in a liquid crystal. Note that 2R~25 nm, and the length of the liquid crystal molecules is about 2-5 nm.

The study of electric transport properties in the liquid crystal 5CB with suspended BaTiO$_3$ nanoparticles is focused on the time transient processes of electric transport caused by the polarization effects and charge accumulation in the samples under different conditions. For this purpose, we carried out measurements of the current versus time in the cells after applying the constant electric voltage across the sample of various magnitudes, as well as time decay of the voltage across the sample in the no-load regime after switching off the power supply. For the sake of comparison, the pure 5CB liquid crystal cells without nanoparticles were also studied.

The main part of electrical scheme for transient characteristics measurements is shown in **Fig. 2**.

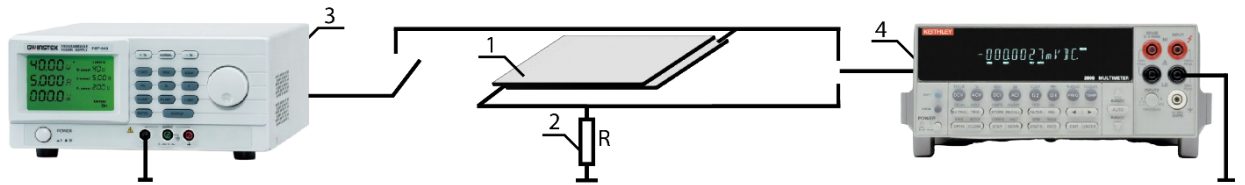

**Fig. 2.** Principal scheme for electrical measurements: *1* – a liquid crystal cell with or without nanoparticles, *2* – a series resistor R for measuring current through the sample; *3* – Power supply "GW Instek PSP-603"; *4* –Voltmeter "Keithley 2000 digital multimeter".

The technique of measurements consists in the following. The voltage of the constant magnitude is applied across the cell and the decaying current through the cell is measured during 120 s. The current is determined from the voltage drop across the series resistor R. The voltage was recorded by the digital multimeter Keithley 2000 conjugated with the computer. After that the power supply is switched off, the switch to the no-load regime occurs and the time decay of the voltage across the sample is recorded. To take off the residual charge from the sample after completing each measurement, the end contacts of the samples are short-cut until the total voltage decreases down to 0.

Shown in **Fig. 3** are time dependencies of the current decay through the samples after applying the constant voltage across them for the suspension of 5CB with 2 concentrations of dispersed $BaTiO_3$ nanoparticles, and pure 5CB (for comparison). For all samples one observes long-lasting transient processes with a duration of several hundred seconds. The total duration of decay is approximately the same in all samples and the curves do not obey a simple exponent law. The difference between the curves is mainly in the magnitude of the current and determined by the nanoparticles concentration.

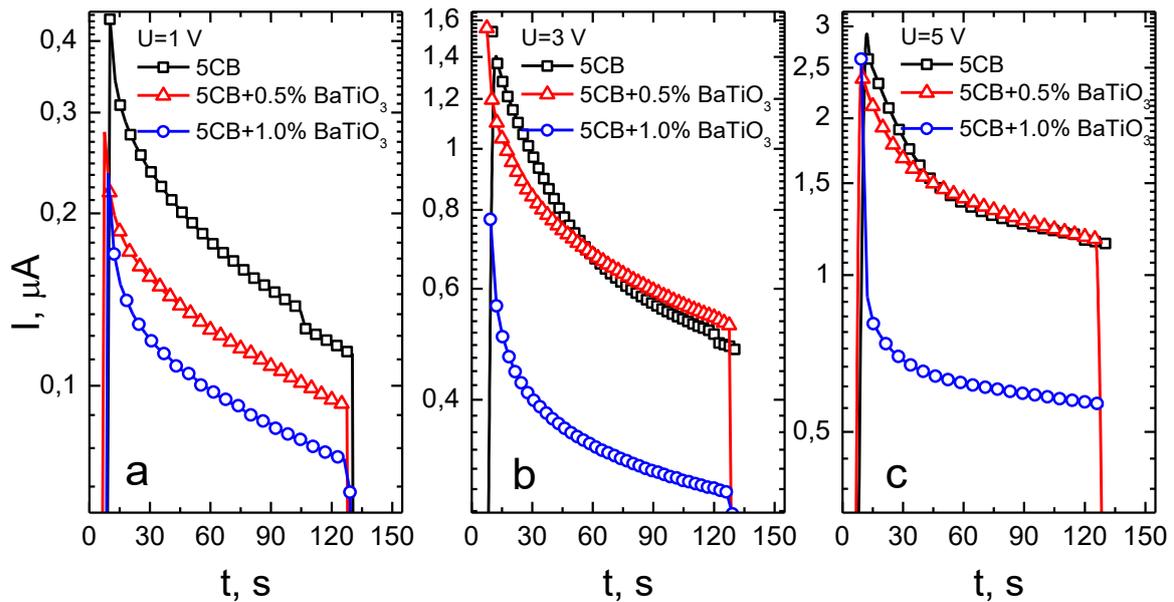

**Fig. 3.** Time dependences of current decay through the cells after applying the constant voltage across them. Applied voltage is 1 V **(a)**, 3 V **(b)** and 5 V **(c)**.

The first set of the curves obtained at the voltage magnitude of 1 V is re-plotted in **Fig. 4** in the coordinates $ln(ln(J_0/J))$ and $ln(t - t_0)$. Here $J_0$ and $t_0$ are the current magnitude and time at the beginning of the current decay curve, respectively. It is seen that all curves are very well fitted by the linear behavior in such coordinates, which evidences them to obey the stretched-exponent law:

$$J = J_0 exp\left[-\left(\frac{t-t_0}{\tau}\right)^\beta\right] + a. \tag{1}$$

We neglect the value of $a$ as compared to the total range of variation of the current $J_0$. The stretched exponent law is characteristic for transient processes in the spatially disordered systems and in our case, it may evidence both of the non-uniform structure of the 5CB suspension and non-uniform distribution of the ferroelectric nanoparticles. From the curves in **Fig. 3** one obtains $\beta$ to be approximately 0.3 – 0.32 and $\tau$ to be of 260 – 300 sec.

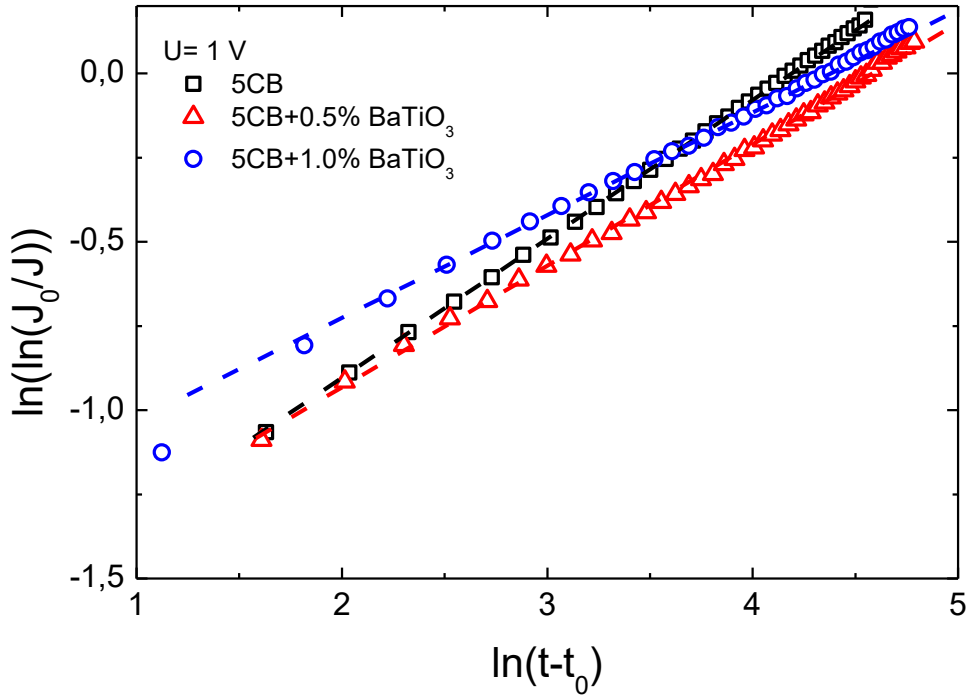

**Fig. 4.** Fitting of the current decay curves by the stretched-exponent law. The fitted curves (dotted lines) and measured points (symbols) are shown for the voltage drop 1 V across the sample. The symbols designating samples correspond to those in **Fig. 3(a)**.

Shown in **Fig. 5** are the time dependencies of the voltage decay on the cells in the no-load regime after switching off the power supply. It is seen that decay does not obey the simple exponent law. Instead, its behavior is more complicated compared to the current decay during the charging process. In general, right after switching off the power supply one observes a sharp decrease in the voltage, which may be related to the rapid polarization process. Then the samples with dispersed ferroelectric nanoparticles and without them manifest somewhat different behavior. The pure 5CB cell comes into approximately single-exponent decay, the cells with suspended nanoparticles demonstrate the decay, which overlaps the quick decaying part and a considerable part of the "next" decay. This behavior may be described by the stretched-exponent law (**Fig. 6**) after which the decay curve deviates to the less steep decay with almost a single-exponent behavior.

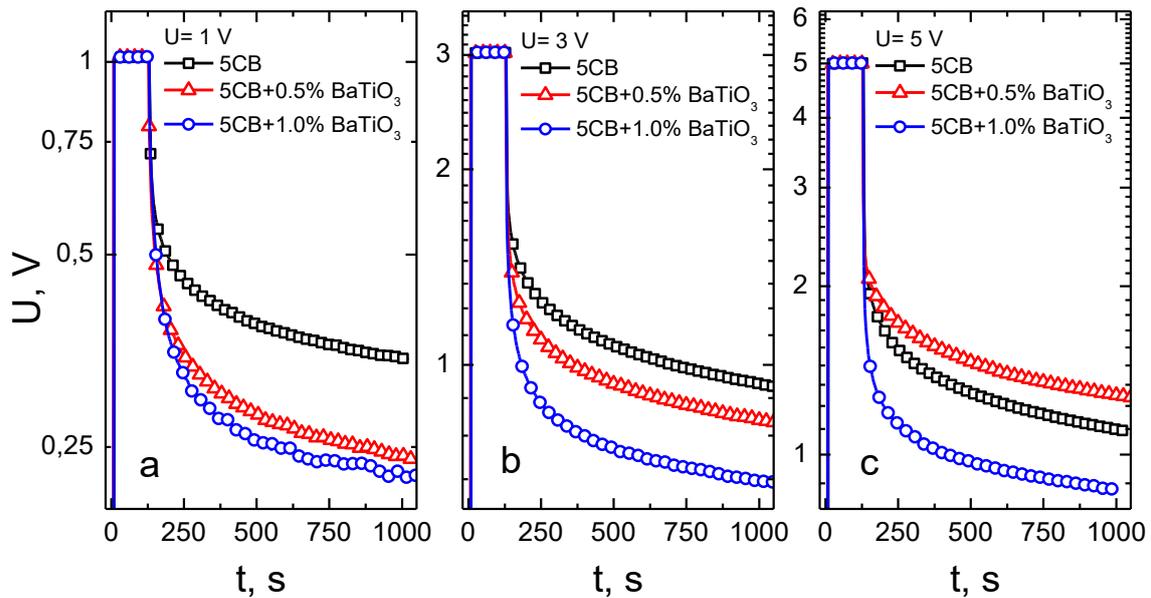

**Fig. 5.** Time dependencies of the voltage decay on the samples in the no-load regime after switching off the power supply. Maximal applied voltage is 1 V **(a)**, 3 V **(b)** and 5 V **(c).**

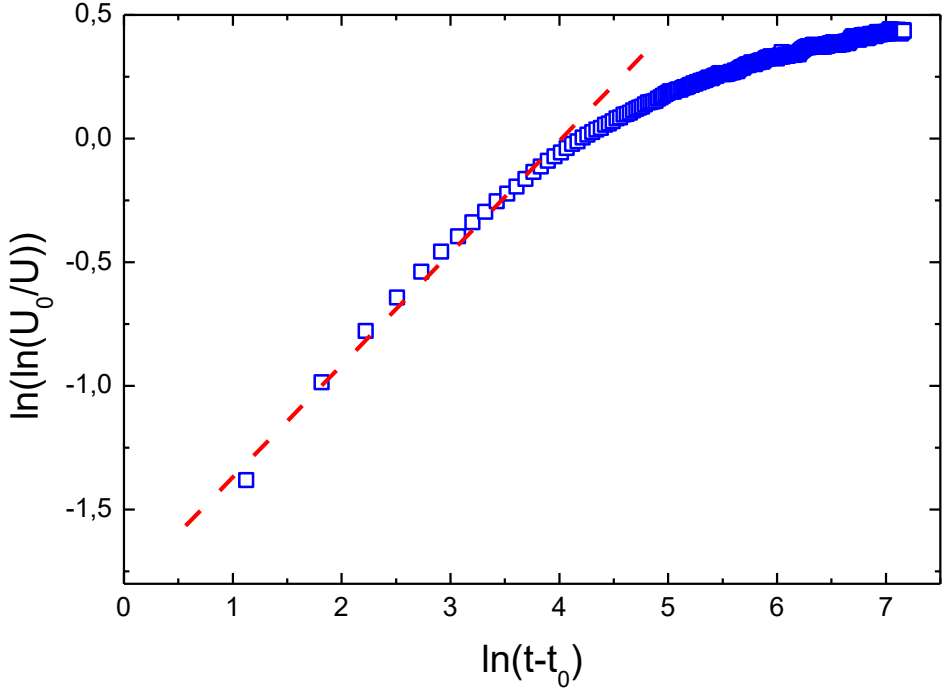

**Fig. 6.** Fitting of the voltage decay curve for the liquid crystal 5CB with 1 wt.% of BaTiO$_3$ nanoparticles by the stretched-exponent law.

## 3. Analysis of theoretical results

To clarify the physical reason for the time dependences of current and voltage, numerical simulation of the electrophysical state of suspensions was carried out by the finite element modelling (FEM) using the phenomenological Landau-Ginsburg-Devonshire (LGD) approach for the calculations of particle polarization and its domain structure dynamics, electrostatic equations for the self-consistent calculations of electric field, and elasticity theory for the modeling of the elastic strains and stresses. Details of the calculations, the system of nonlinear LGD equations with boundary conditions, and material parameters of BaTiO$_3$ are given in Ref. [11].

The simulation assumes that the surface of a polarized ferroelectric nanoparticle accumulates an ionic-electronic charge that partially or completely screens the electric polarization outside the particle. The density of this charge depends on the state of the polarization at the nanoparticle surface and on the properties of the surrounding liquid medium [20]. The ionic-electronic charge density determines the effective screening length λ of the spontaneous polarization at the nanoparticle surface.

The Stephenson-Highland model [21] describes the relationship between surface charge density $\sigma_S[\delta\phi]$ and electric potential excess $\delta\phi$ at the nanoparticle surface. This model accounts for the coverages of positive and negative surface charges (e.g., ions, anions, and electrons) in a self-consistent manner. The corresponding Langmuir adsorption isotherm is given by expression [22]:

$$\sigma_S[\delta\phi] \cong \sum_i \frac{eZ_i}{A_i}\left(1 + a_i^{-1}\exp\left[\frac{\Delta G_i + eZ_i\delta\phi}{k_B T}\right]\right)^{-1}, \qquad (2a)$$

where $e$ is the electron charge, $Z_i$ is the ionization number of the adsorbed ions, $a_i$ is the dimensionless chemical activity of the ions in the environment (as a rule $0 \leq a_i \leq 1$), $T$ is the absolute temperature, $A_i$ is the area per surface site for the adsorbed ion, and $\Delta G_i$ are the formation energies of the positive (subscript $i = 1$) and negative (subscript $i = 2$) surface charges (e.g., ions and/or electrons) at normal conditions.

Linearization of Eq.(2a) with respect to the electrostatic potential gives the expression for the effective screening length $\lambda$ associated with the adsorbed charges [20]:

$$\frac{1}{\lambda} \approx \sum_i \frac{(eZ_i)^2 a_i \exp\left[\frac{\Delta G_i}{k_B T}\right]}{\varepsilon_0 k_B T A_i\left(a_i + \exp\left[\frac{\Delta G_i}{k_B T}\right]\right)^2} \qquad (2b)$$

For relatively large λ (more than 5 - 10 nm) the polarization screening is weak, and the nanoparticle is in the non-polar paraelectric phase induced by its small size (~25 nm). With decreasing λ (from 10 to 1 nm) the screening increases and the spontaneous polarization of the nanoparticle preserves. However, the screening is not strong enough to maintain the single-domain polarization state of the nanoparticle. Since a barium titanate is a multiaxial ferroelectric, the polarization of the nanoparticles can rotate, forming vortex-like structures that significantly reduce the depolarization field inside the particle and the scattering field outside them [23]. A since-domain state of the polarization inside the particle is energetically preferable at very small λ (much less than 1 angstrom). "Perfect" screening (λ→0) minimizes the indirect electrostatic interaction between the nanoparticles via the liquid crystal and the direct interaction of the particles with the liquid crystal.

Thus, the analysis of the polarization state of a nanoparticle at large and small λ seems urgent to explain the experimental observations. Therefore, the FEM was carried out for λ within 0.01 – 10 nm, which provides a significant influence of the ionic-electronic charge on the ferroelectric polarization of the particle ("ferro-ionic" coupling) and does not exclude long-range electrical interactions between the particles with increasing λ. We also assume that λ significantly increases with decreasing nanoparticle concentration, since the immersion of nanoparticles covered by surfactant leads to the appearance of additional ionic-electronic charge in the dielectric liquid crystal 5CB.

The dependence of the charge density on the electrodes, between which there is a cell with a liquid crystal 5CB with densely packed 24-nm spherical $BaTiO_3$ nanoparticles (particle concentration is more than 50 vol.% in the case) is shown in **Fig. 7(a)**. The bottom part of the figure shows the distributions of spontaneous polarization $P_S$ inside the $BaTiO_3$ nanoparticle and the electric potential $\varphi$ inside and around the particle in the states "+1" and "-1", which are indicated on the charge loop. The nanoparticles are divided into domains, the direction of polarization inside which is determined by the direction of voltage sweep. The polarization of the domains in the states "+1" and "-1", which correspond to zero external voltage, is opposite. The interaction of the polarization of the domains with the liquid crystal and with other nanoparticles leads to a specific potential distribution, which is also inverted in the states "+1" and "-1". Since the single-domain state is not observed at low voltages, the hysteresis loop of the charge stored by the liquid crystal cell is narrow and inclined, and the maximal spontaneous polarization in the "+1" and "-1" states (~0.06 C/m$^2$) is approximately 4 times smaller than the polarization of a bulk barium titanate single-crystal at room temperature (~0.25 C/m$^2$).

**Figures 7(b)-(d)** show the charge density dependences on the electrodes covering the cell with is a suspension of liquid crystal with 10 vol.%, 1 vol.% and 0.1 vol.% of BaTiO$_3$ nanoparticles, respectively. The charge density dependences are characterized by a very weak hysteresis, mainly associated with the polarization dynamics of the liquid crystal organic molecules. An increase in the concentration of BaTiO$_3$ nanoparticles leads to the appearance of small additional "loops" or "self-intersections" located symmetrically at the ends of the main loop of the voltage-charge characteristic. The nanoparticles themselves remain multi-domain at low voltages (as shown in the insets), but the magnitude of the polarization in the domains and the contrast decrease with increase in $\lambda$. When $\lambda$ is above 5 nm, the surface charge does not play any significant role in the screening of polarization in the 24 nm BaTiO$_3$ particles.

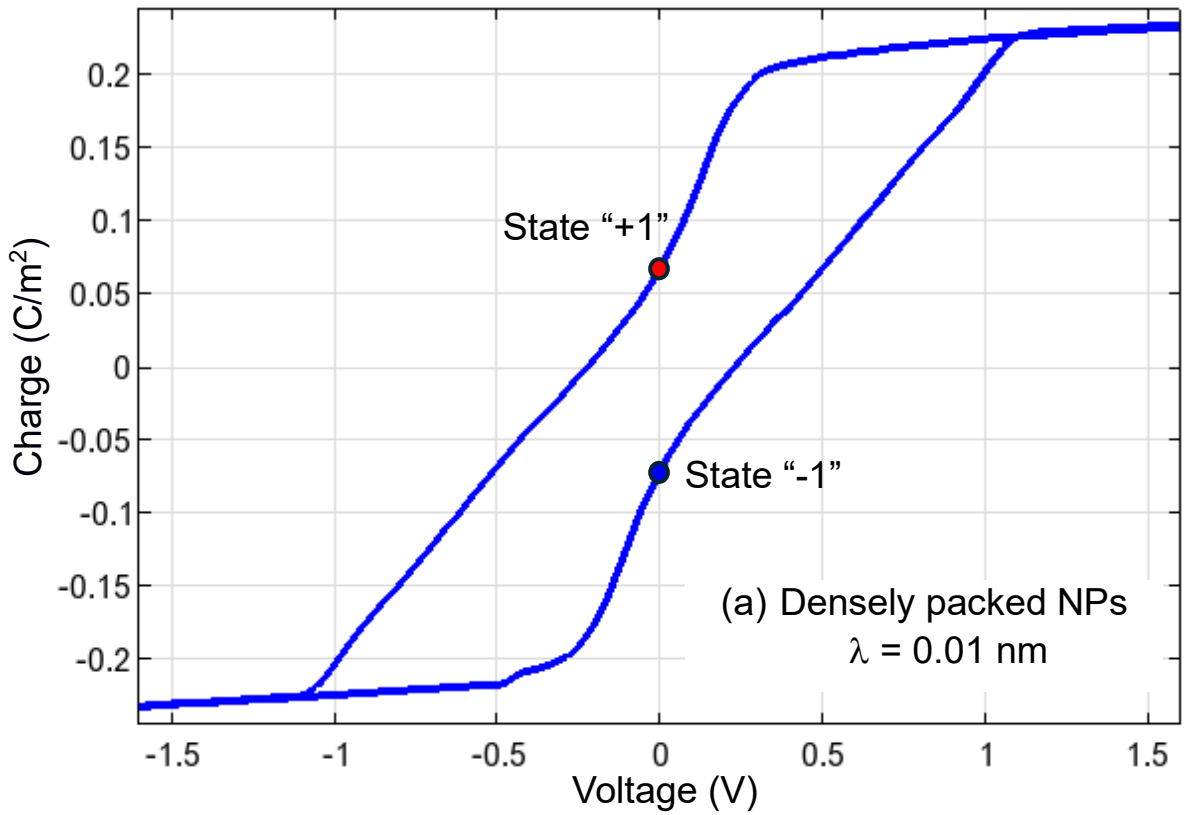

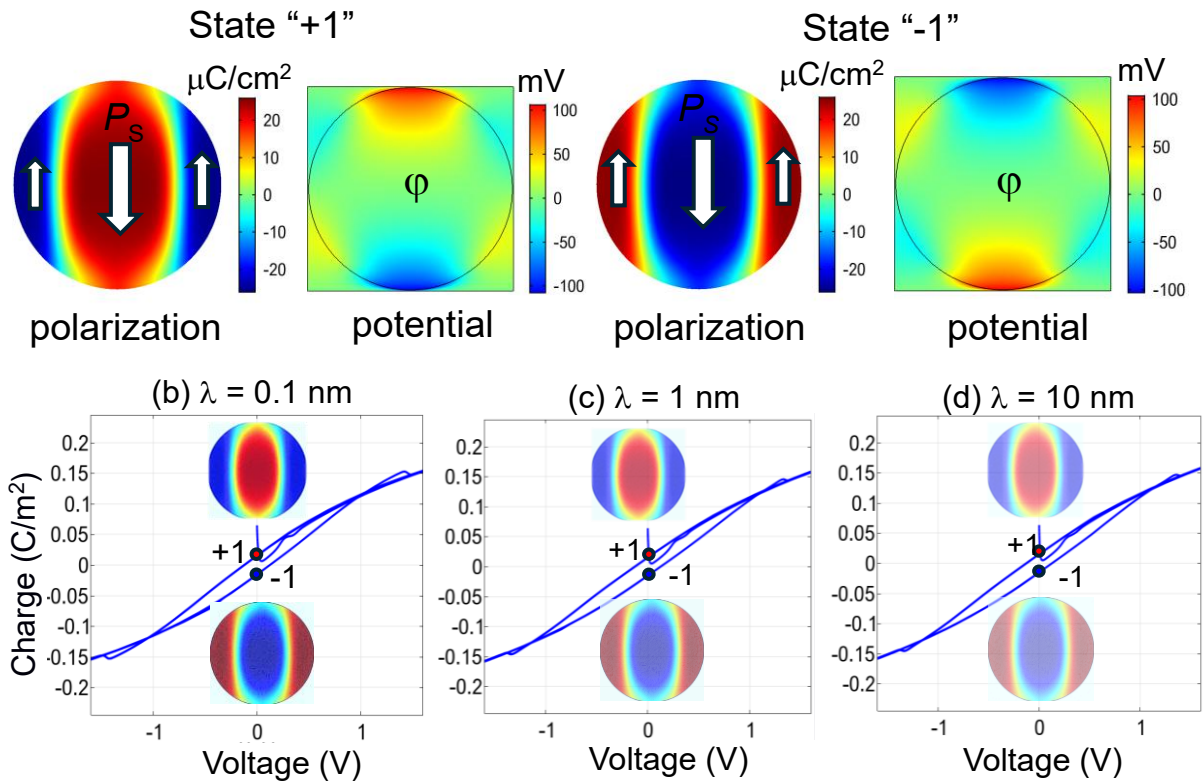

**Fig.7. (a)** Dependence of the charge density on the electrodes, between which there is a cell with a liquid crystal 5CB, where spherical BaTiO$_3$ nanoparticles are densely packed. The electric polarization of the particles is screened by the ionic-electronic charge adsorbed at their surface. The bottom line are the distributions of the spontaneous electric polarization

$P_S$ and the electric potential $\varphi$ of the BaTiO$_3$ nanoparticle in the states "+1" and "-1", which are indicated on the charge loop. Parts **(b)-(d)** show the dependences of the charge density on the electrodes, between which there is a liquid crystal suspension, which contains about 10 vol.% **(b)**, 1 vol.% **(c)** and 0.1 vol.% **(d)** of BaTiO$_3$ nanoparticles, respectively. The FEM was carried out at room temperature, the radius of the nanoparticles $R$ =12 nm, the screening length $\lambda = 0.01$ nm (a), $\lambda = 0.1$ nm (b), $\lambda = 1$ nm (c) and $\lambda = 10$ nm (d).

Thus, the FEM results confirmed the leading role of the ionic-electronic screening charges, which screen dynamically the spontaneous polarization outside the particle. When an electric field is applied to a suspension of nanoparticles, it releases some of the screening charge (mainly due to the field-induced changes of the nanoparticle polarization), leading to decrease of the current and voltage decay times of the cell with liquid crystal and nanoparticles (in comparison with the liquid crystal cell without nanoparticles).

Note that the conclusion is valid at relatively small voltages applied to the cell with small concentration of BaTiO$_3$ nanoparticles, which agrees with the time dependences shown by the black, red and blue curves in **Fig. 3(a)**, **Fig. 5(a)** and **5(b)**. Under the voltage increase the nonlinear process may dominate leading to the nonmonotonic dependence of the decay time on the concentration of nanoparticles (e.g., see the sequence of black, red and blue curves in **Fig. 5(c)**).

At the same time, the higher the concentration of nanoparticles in the liquid crystal, the greater the decrease in the decay time may be expected. However, one cannot expect a linear dependence of the decay time on the concentration of nanoparticles in the liquid crystal, since the process of surface adsorption of the ionic-electronic charge and its interaction with the polarization of the particles is strongly nonlinear. The indirect dipole-dipole interaction of nanoparticles in the liquid crystal is also nonlinear. With an increase in the concentration of particles (more than 10 %), saturation of the dependence of the decay time on the concentration of nanoparticles is possible.

## 4. Conclusions

The work analyses the electrophysical properties of the cells filled with pure nematic liquid crystal 5CB and a liquid crystal suspension containing 0.5 wt.% and 1 wt.% of $BaTiO_3$ nanoparticles.

Specifically, we analyze the time dependences of the current flowing through the sample at fixed values of the applied electric voltage and voltage decay across the sample in the no-load mode after switching off. The time dependences of the current and voltage show a retarding decay rate. For samples with $BaTiO_3$ nanoparticles, a retarding decay time rate is characteristic. Possible physical reasons for this are the influence of screening charges, which cover ferroelectric nanoparticles and are polarized in the external electric field, and slow ionic transport in the liquid crystal.

The FEM of the polarization distribution, domain structure dynamics, and charge state of nanoparticles, performed using LGD approach, confirmed the leading role of the screening charges, because the surface of a polarized ferroelectric nanoparticle adsorbs an ionic-electronic charge that partially screens its electric polarization in single-domain and/or poly-domain states. When an electric field is applied to the nanoparticles, it releases some of the screening charge due to a change in their polarization, leading to elongation of the decay time of current and voltage on a cell with liquid crystal and nanoparticles.

**Acknowledgements.** The work of Y.M.G., O.S.P., D.O.S., E.A.E. and A.N.M. is funded by the National Science Foundation of Ukraine (grants No. 2023.03/0132 "Multiple degenerated metastable states of spontaneous polarization in nanoferroics: theory, experiment and prospects for digital nanoelectronics" and No. 2023.03/0127 "Silicon-compatible ferroelectric nanocomposites for electronics and sensorics"). The work of V.V.V. and V.M.P. is funded by the target program of the NAS of Ukraine, project No. 5.8/25-P

"Energy-saving and environmentally friendly nanoscale ferroics for the development of sensors, nanoelectronics and spintronics. The work of O.V.B. is funded by the NAS of Ukraine, grant No. 07/01-2025(6) "Nano-sized multiferroics with improved magnetocaloric properties". S.E.I. and A.N.M. also acknowledge the the NATO Science for Peace and Security Programme under grant SPS G5980 "FRAPCOM" for sponsoring samples preparation, characterization and results analysis.

**Author's contribution.** Electrophysical measurements and analysis of experimental results were carried out by Y.M.G., O.S.P., V.V.V. and V.M.P. V.V.V. wrote the experimental part of the work. I.A.G. prepared the suspensions, S.E.I. synthesized the nanoparticles. D.O.S. prepared the cells for measurements. E.A.E. wrote the codes for FEM and tested them together with O.V.B. A.N.M. stated the problem and analyzed theoretical results, wrote the theoretical part of the work, introduction and conclusions (together with V.V.V.). All co-authors participated in the discussion of obtained results and improving the text of the article.